\documentclass[onecolumn]{revtex4}
\usepackage[]{graphicx}
\usepackage{epstopdf}
\usepackage{xcolor}
\usepackage[]{textcomp}
\usepackage{amsmath,amssymb,times}
\usepackage[T1]{fontenc}
\usepackage{mathptmx}

\begin{document}
\title{Feasibility study towards comparison of the \emph{g$^{(2)}$}(0) measurement in the visible range}

\author{E. Moreva$^1$, P. Traina$^1$\footnote{e-mail: p.traina@inrim.it}, R. A. Kirkwood$^2$, M. L\'{o}pez$^3$, G. Brida$^1$, M. Gramegna$^1$, I. Ruo-Berchera$^1$, J. Forneris$^{4,5}$, S. Ditalia Tchernij$^{6,4}$, P. Olivero$^{6,4,1}$, C. J. Chunnilall$^2$, S. K\"{u}ck$^3$, M. Genovese$^1$, I. P. Degiovanni$^1$}

\affiliation{$^1$Istituto Nazionale di Ricerca Metrologica (INRIM) Torino, Italy}

\affiliation{$^2$National Physical Laboratory (NPL), Teddington, UK}

\affiliation{$^3$Physikalisch-Technische Bundesanstalt (PTB), Braunschweig and Berlin, Germany}

\affiliation{$^4$Istituto Nazionale di Fisica Nucleare (INFN) Sez. Torino, Torino, Italy}

\affiliation{$^5$Ru\dj{}er Bo\v{s}kovi\'{c} Institute - Bijenicka 54, P.O.Box180, 10002, Zagreb, Croatia}

\affiliation{$^6$Physics Department and NIS inter-departmental centre - University of Torino, Torino, Italy}

\keywords{Single-photon source, photon statistics, quantum metrology}

\begin{abstract}
This work reports on the pilot study, performed by INRIM, NPL and PTB, on the measurement of the $g^{(2)}(0)$ parameter in the visible spectral range of a test single-photon source  based on a colour centre in diamond. The development of single-photon sources is of great interest to the metrology community as well as the burgeoning quantum technologies industry. Measurement of the $g^{(2)}(0)$ parameter plays a vital role in characterising and understanding single-photon emission. This comparison has been conducted by each partner individually using their own equipment at INRIM laboratories, which were responsible for the operation of the source. 
\end{abstract}
\maketitle
\section{Introduction}
Single-photon sources (SPSs) \cite{mig,genjo,sps}, i. e. sources that are able to produce 
 single photons on demand, can prove to be key elements for the development of quantum optical technologies. They will also be essential for providing metrological support   for the development and commercialisation of these technologies, as well as for radiometry and photometry at the single-photon level. 
 SPSs based on different physical systems (Parametric down-conversion \cite{sramelow,sorgentina,krapick,fortsch,montaut,oxborrow,eiseman,sorgentina2}, quantum dots \cite{dots,arita}, trapped ions \cite{ion}, molecules \cite{mol} and colour centres in diamond \cite{dia1,dia2,dia3,dia4,dia5,dia7,dia6,el,tin,he}) and single-photon sensitive detectors \cite{exc,tes,itz} and cameras \cite{spadcam} are  widely available today as well as more complex equipment such as quantum key distribution systems \cite{idq,tosh}. Despite several recent dedicated studies \cite{aigar,bea}, a standardized methodology for the characterization of SPSs  has not emerged.  

The typical parameter employed to test the properties of a SPS is the second order correlation function (or Glauber function) defined as 
 \begin{equation}
 g^{(2)}(\tau=0)=\frac{\langle I(t)I(t+\tau)\rangle}{\langle I(t)\rangle\langle I(t+\tau)\rangle}\Biggr\rvert_{\tau=0},
 \end{equation}
 
where $I(t)$ is the intensity of the optical field. In the regime of low photon flux, this parameter  has been shown to be substantially equivalent to the parameter $\alpha$ introduced by Grangier et al. \cite{gran}, which is experimentally measured as the ratio between  the coincidence probability at the output of a Hanbury Brown and Twiss (HBT) interferometer  \cite{hbt}, typically implemented by a 50:50 beam-splitter connected to two non photon-number-resolving detectors, and the product of the click probabilities at the two detectors, i. e.:
\begin{equation}\label{eq:alfa}
g^{(2)}(\tau=0)\approx \alpha =\frac{P_{C}}{P_AP_B},
\end{equation}
where $P_{C}$, $P_A$,  $P_B$ are, respectively,  the coincidence and click probabilities at the outputs $A$, $B$ of  an HBT interferometer.This identity holds strictly for very low value of the click-probabilities $P_A$ and $P_B$ (namely much less than 0.1), i. e. for very faint SPSs, while it is only approximately verified for brighter sources. Due to the equivalence between $g^{(2)}(0)$ and $\alpha$ in the regime typical of quantum optics experiments, all experimental measurements of $g^{(2)}(0)$ in the relevant literature are actually measurements of $\alpha$, since the two parameters are used substantially without distinction in this community.

This work presents a systematic study of the $\alpha$ measurement for a SPS in pulsed regime, with the purpose of developing a measurement procedure and an analysis of the uncertainty to provide an unbiased value of the measurand which is independent of the experimental apparatus used and ultimately producing an estimate unaffected by the non-ideal behavior of the physical systems. 
Consensus on such a procedure would produce great benefits for the metrology community, enabling the development of SPS characterization techniques that are robust enough for practical measurement services. The results reported in this work were obtained during a pilot study 
performed by INRIM,  NPL and PTB. This is a precursor to organising of an international comparison on the $g^{(2)}(0)$ measurement, which would pave the way for the realization of a mutual recognition agreement on the calibration of  key elements for the forthcoming quantum technologies, such as SPSs and single-photon detectors. This comparison was hosted at INRIM from October 16$^{th}$ to October 29$^{th}$ 2017 and was composed of two joint measurements of  $\alpha$ on the same emitter: one performed by INRIM and PTB and the other one by INRIM and NPL. This procedure was adopted because it allows  the results of two measuring devices operating simultaneously to be compared. Measurements on the same source at different times can yield slightly different results, since the imperfectly reproducible alignment of the source can lead to a different amount of noise coupled to the detection system. An SPS based on a Nitrogen-Vacancy centre excited in the pulsed regime, emitting single photons in the spectral range from 650 nm to 750 nm  was used as a source.  

An analogous effort to establish a proper procedure
for the measurement of the $g^{(2)}$ function of a telecom heralded SPS (in continuous regime) can be found in \cite{gino}.

\section{Measurement Technique}
With regards to Eq. \ref{eq:alfa}, probabilities $P_C$, $P_A$,  $P_B$ are estimated as the ratio between the total number of the corresponding events  versus the number of excitation pulses during the experiment, i. e. $P_x = N_x/(R*t_{acq}) (x=C, A, B)$, where $R$ is the excitation rate and $t_{acq }$ is the total acquisition time.
The value of the measurand is independent from the total efficiencies ($\eta_A$, $\eta_B$) of individual channels (including detection and coupling efficiency), optical losses and splitting ratio since
\begin{equation}
\alpha =\frac{\eta_A \eta_B P_C}{\eta_A P_A\eta_B P_B}=\frac{P_C}{P_AP_B}.
\end{equation}
The value of the parameter from the experimental data, corrected for the contribution of the background coincidences (due, for example, to stray light or residual excitation light), can be estimated as follows:
\begin{equation}
\alpha =\frac{P_C-P_{Cbg}}{(P_A-P_{Abg})(P_B-P_{Bbg})},
\end{equation}
where $P_{Cbg}$, $P_{Abg}$,  $P_{Bbg}$ are, respectively, the coincidence and click probabilities of background photons, calculated analogously to their counterparts $P_{C}$, $P_A$,  $P_B$.

\begin{figure}[!ht]
\centering
\includegraphics[scale=0.5]{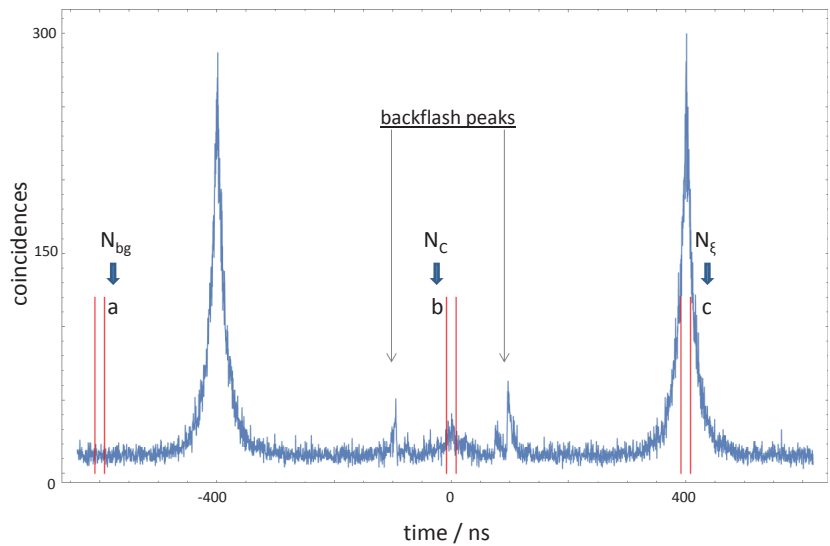}
\caption{Typical chronogram of the coincidence events registered with an HBT interferometer. The three identical highlighted time intervals are used to estimate, respectively, background coinicidences (a), true coincidences (b) and accidental coincidences (c).}
\label{fig:isto}
\end{figure}

Fig. \ref{fig:isto} shows the typical chronogram of the behaviour of a pulsed SPS obtained by sampling the coincidence events at the two outputs of an HBT interferometer. The coincidence probability has been estimated as the ratio between the total number of events in the chronogram falling in a proper temporal window $w$ around the central peak (showing antibunching, i.e. the "b" interval in Fig. 1) and the total number of excitation pulses occurring in the acquisition time. The product $P_A P_B$, corresponding to the probability of accidental coincidences, has been evaluated by integrating the events occurring in an equal interval around the subsequent peak ("c" interval in Fig. 1) not showing antibunching (always divided by the number of pulses). In fact, those coincidence events (amounting to $N_\xi$) are related to independent events (coincidences between single photons emitted after two subsequent laser pulses and detected by detector A and detector B respectively) and thus  $P_\xi - P_{bg} =(N_\xi- N_{bg})/(R*t_{acq})$. 

The parameter to be extimated is thus:
\begin{equation}
\alpha_{exp}=\frac{N_C-N_{bg}}{N_\xi-N_{bg}},
\end{equation}
where ($N_i$ being the coincidence events sampled in the $i$-th channel)
\begin{eqnarray}
N_C &=& \sum_{i=-k_w/2}^{k_w/2}{N_i},\\
N_\xi &=& \sum_{i=T-k_w/2}^{T+k_w/2}{N_i},
\end{eqnarray}
$k_w$ is the number of bins corresponding to the chosen coincidence window $w$, $N_{bg}$ is the estimated background due to spurious coincidences (the number of events in the "a" interval in Fig 1) and $T$ is the excitation period (expressed in bins).

In Fig. \ref{fig:isto} two backflash peaks \cite{bf,bf2,bf3} can be observed on either side of the central peak. Those are due to secondary photon emission that arises from the avalanche of charge carriers that occurs in one of the two detectors in the HBT as a photon is absorbed and that are afterwards detected from the other detectors. To avoid overestimating $\alpha$, these  peaks must not be included in the coincidence window. 

The presence of the backflash peaks prevented us to estimate $P_A$, $P_B$ directly from tha counts of the two detectors, since we were forced to consider a coincidence window smaller than the NV-center emission time window (of the order of tens of nanoseconds, i.e. at least three NV lifetimes). For this reason we estimated $P_A*P_B$ consistently with the coincidences measured at time 0. 
The probability of observing a coincidence in the autocorrelation window around the peak at 400 ns can be underestimated by the presence of the coincidence counts between 0 and 400 ns, because of detectors and electronics dead-time. Due to the extremely low level of counts in this interval, we have estimated that this correction is negligible within the declared probability uncertainty.

\section{Measurement facility}
\begin{figure}[!ht]
\centering
\includegraphics[scale=0.53]{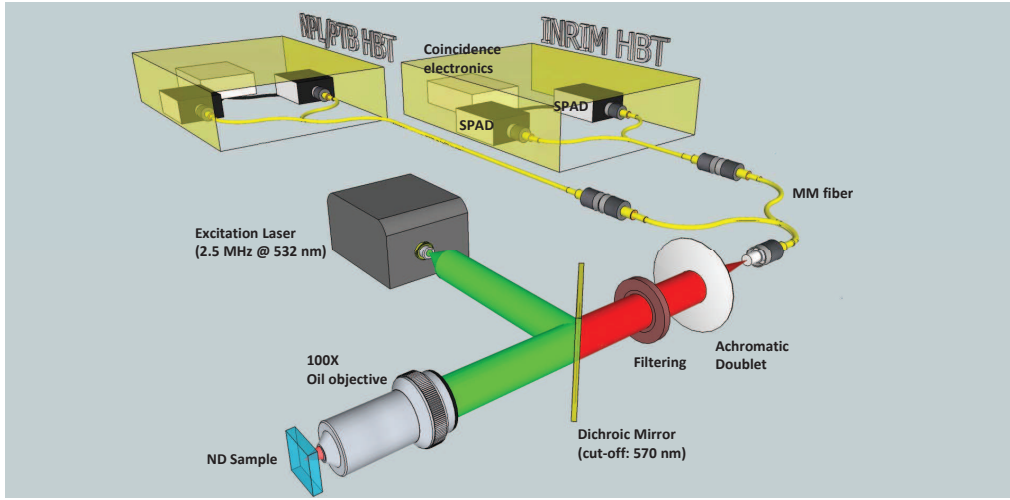}
\caption{Scheme of the experimental setup: the output of a laser-scanning confocal microscope is split by a 50:50 beam-splitter and directed to two independent HBTs measurement systems performing the comparison. One of the measurement devices is held by the host institution (INRIM) while the other one is used, in turn, by the other two partners (NPL, PTB)}
\label{fig:setup}
\end{figure}

Fig. \ref{fig:setup} shows the experimental setup: a laser-scanning confocal microscope whose signal is split by a 50:50 beam-splitter connected to two measurement devices, i. e. two single-photon sensitive HBT interferometers. Note that, according to the model described in Sec. II, the value of $\alpha$ measured by the two HBT interferometers is independent of optical losses and splitting ratio at the beam-splitter. The excitation light, produced by a pulsed laser (48 ps FWHM, 560 pJ per pulse) emitting  at 532 nm with a repetition rate $R=$ 2.5 MHz was focused by a 100$\times$ oil-immersion objective on the nano-diamond (ND) sample hosting an SPS based on a single NV center of negative charge, with emission in a broad spectral band starting approximately at 630 nm and ending at 750 nm ($\lambda_{ZPL}$ = 638 nm) \cite{dia2}. The optical filters used were a notch filter at 532 nm and two long-pass filters (FEL600 and FEL650). The photoluminescence signal (PL) , thus occurring in a  650 nm - 750 nm spectral range,  was collected by a multimode fibre and split by a 50:50 beam-splitter (BS). As stated above, each end of the BS was connected to a separate HBT setup used for the joint measurement. In particular:
\begin{itemize}
\item{The INRiM facility was composed of a fused 50:50 fibre beam-splitter connected to two Excelitas SPCM-AQR-14-FC Single-Photon Avalanche Detectors (SPADs). Single and coincidence counts were sampled via ID Quantique {\em ID800} time-to-digital converter (60 ps time resolution).}\\
\item{The NPL facility was composed of a fused 50\textcolor{red}{:}50 fibre beam-splitter connected to two Perkin-Elmer SPCM-AQR-14-FC Single-Photon Avalanche Detectors (SPADs). Coincidence counts were sampled via PicoQuant {\em HydraHarp 400} multichannel picosecond event timer (1 ps time resolution).}\\
\item{The PTB facility was composed of a fused 50\textcolor{red}{:}50 fibre beam-splitter connected to two Excelitas SPCM-AQR-14-FC Single-Photon Avalanche Detectors (SPADs). Single and coincidence counts were sampled via PicoQuant {\em HydraHarp 300} multichannel picosecond event timer (4 ps time resolution).}  
\end{itemize}

The detailed description of the sample fabrication and preparation is reported elsewhere \cite{quasiteng}.

\section{Results}
\begin{figure}[ht]
\centering
\begin{tabular}{cc}
\includegraphics[scale=0.85]{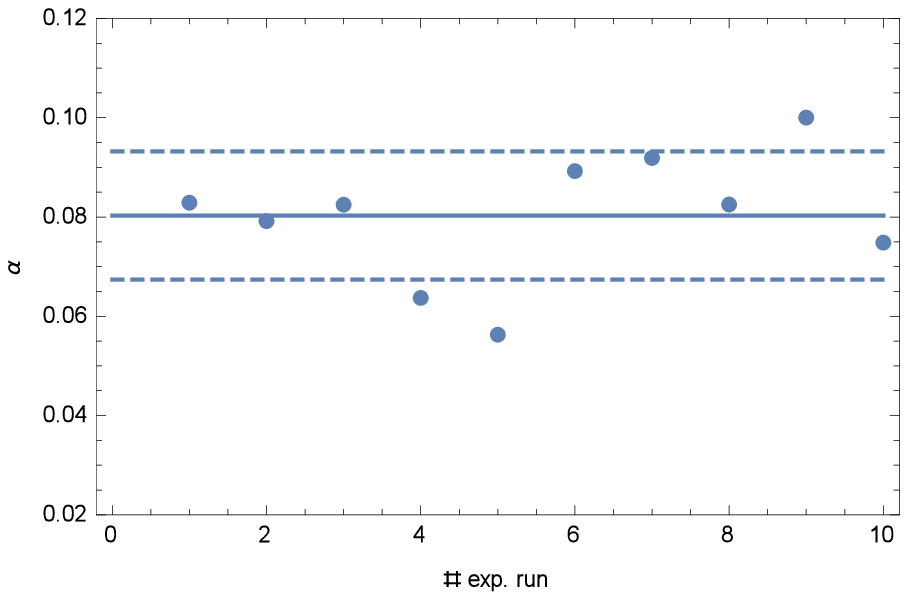}&\includegraphics[scale=0.85]{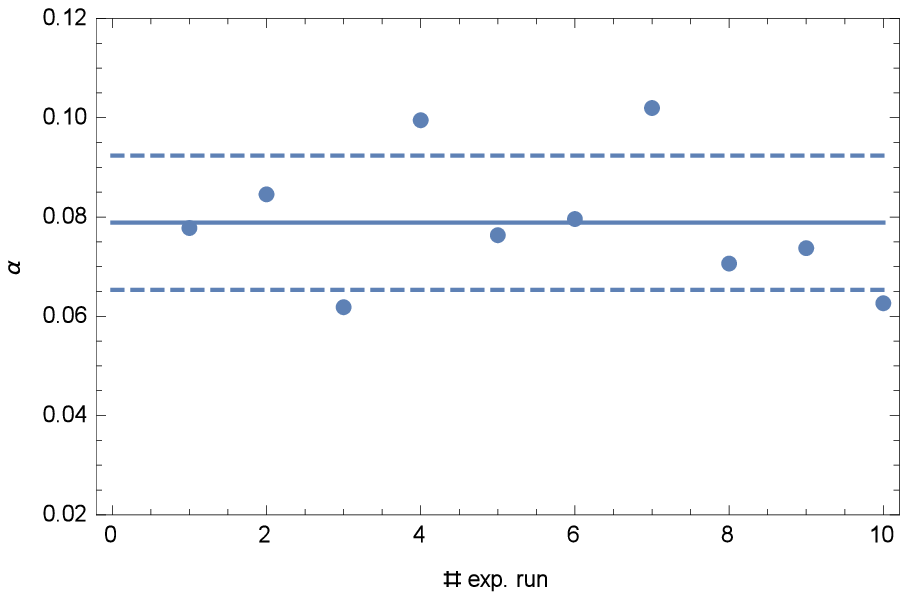}\\
\end{tabular}
\caption{Distribution of the individual measurements 
 performed by INRIM (left) and PTB (right) in joint measurement.}
\label{fig:g2_summ_11}
\end{figure}

\begin{figure}[ht]
\centering
\begin{tabular}{cc}
\includegraphics[scale=0.85]{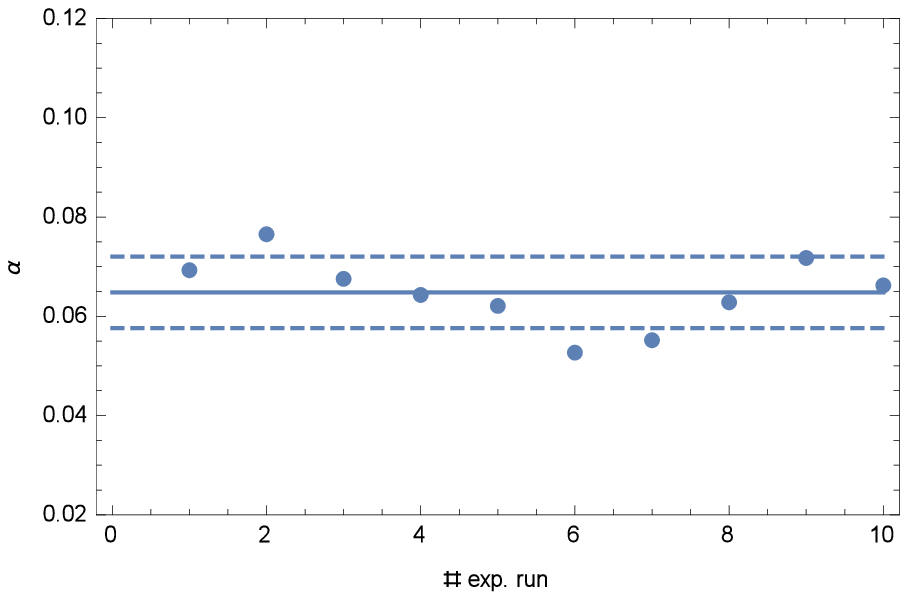}&\includegraphics[scale=0.85]{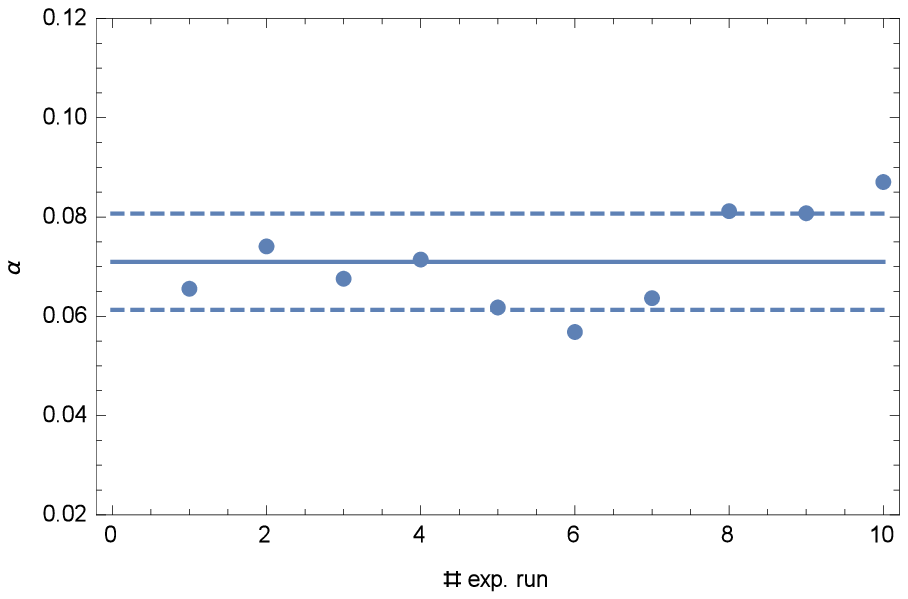}
\end{tabular}
\caption{Distribution of the individual measurements 
 performed by INRIM (left)  and NPL (right) in joint measurement.}
\label{fig:g2_summ_13}
\end{figure}

 Each measurement consisted of 10 runs each of 500 s acquisition time. The total coupling rate,  accounting for limited SPS quantum efficiency, collection angle, optical losses and detection efficiency (excluding the splitting ratio of the detector-tree), has been estimated as the ratio between the counting rate of the detectors (summing over all four of them) and the excitation rate, yielding  $\eta_{TOT}$ = (1.76 $\pm$ 0.01)$\%$. The coincidence window $w$ considered for evaluating the reported $\alpha$ was $w=$ 16 ns. By repeating the analysis for different temporal widths $w$ it was observed that the results were consistent as long as the backflash peaks were not included in the coincidence window (see Fig. 5). Figures \ref{fig:g2_summ_11}, \ref{fig:g2_summ_13}  show the distributions of the $\alpha_{exp}$ values measured by each partner; the continuous line indicates the mean value 
and the dashed lines draw a 1-$\sigma$ confidency band around the mean value. Tables I-IV report the uncertainty budgets associated with the measurements. The summary of the results of the joint measurement is presented in Table \ref{tab:sum}. We observe that individual measurement sessions (INRIM/NPL and INRIM/PTB) yield results that are extremely consistent. Mechanical instability in the coupling of the source may be the reason why the two sessions are not perfectly in agreement and the results of the INRIM/PTB joint measurements yield a slightly higher $\alpha$ value (as well as greater associated uncertainty) with respect to the INRIM/NPL ones. In fact, the agreement in the INRIM/NPL measurements is better than indicated by Fig. \ref{fig:g2_summ_13} and the calculations, since the NPL measurements took longer than the INRIM measurements, the last two NPL measurements being performed after INRIM had completed its measurements. However,  
all values are compatible within the uncertainty ($k$=2).  The uncertainties on the results of the measurements have been calculated as combined standard uncertainties for correlated input parameters $N_x$ ($x$ $=C$, $\xi$, $bg$) according to the formula \cite{GUM}:

\begin{widetext}
\begin{equation}
u_c(\alpha_{exp})=\sqrt{\sum_x \biggl(\frac{\partial \alpha_{exp}}{\partial N_x}\biggr)^2u(N_x)^2+2\sum_{x,y}\rho_{xy} \biggl(\frac{\partial \alpha_{exp}}{\partial N_x}\biggr)\biggl(\frac{\partial \alpha_{exp}}{\partial N_y}\biggr)u(N_x)u(N_y)},
\end{equation}
\end{widetext}

where the correlation coefficient $\rho_{xy}$ is defined as

\begin{equation}
\rho_{xy}=\frac{\langle N_x N_y\rangle-\langle N_x\rangle \langle N_y\rangle}{u(N_x)u(N_y)}.
\end{equation}

\begin{table}[!h]
\begin{tabular}{|c|c|c|c|c|}
\hline
Quantity &	Value &	Standard unc. &	Sens. Coeff. &	Unc. contribution\\
\hline
\hline
$N_C$ &	$1000$ &	$70$ &	$1.5 \ast 10^{-4}$ &	$1 \ast 10^{-2}$\\
\hline
$N_\xi$	 & $7400$ &	$900$ &	$-9 \ast 10^{-6}$	 & $-1 \ast 10^{-2}$\\
\hline
$N_{BG}$	& $560$ & 	$30$ &	$-1.4 \ast 10^{-4}$ &	$-3 \ast 10^{-3}$\\
\hline
$\alpha_{exp}$ &	$0.065$ & & &			$0.005$\\
\hline
\end{tabular}
 \caption{Uncertainty Budget ($k$=2) associated with INRIM (in joint measurement with NPL)}
 \label{tab:25inrim}
\end{table}

\begin{table}[!h]
\begin{tabular}{|c|c|c|c|c|}
\hline
Quantity &	Value &	Standard unc. &	Sens. Coeff. &	Unc. contribution\\
\hline
\hline
$N_C$ &	$900$ &	$200$ &	$2 \ast 10^{-4}$ &	$2 \ast 10^{-2}$\\
\hline
$N_\xi$	 & $6000$ &	$2000$ &	$-1 \ast 10^{-5}$	 & $-2 \ast 10^{-2}$\\
\hline
$N_{BG}$	& $540$ & 	$50$ &	$-2 \ast 10^{-4}$ &	$-7 \ast 10^{-3}$\\
\hline
$\alpha_{exp}$ &	$0.068$ & & &			$0.005$\\
\hline
\end{tabular}
 \caption{Uncertainty Budget ($k$=2) associated with NPL measurement (in joint measurement with INRIM)}
 \label{tab:25npl}
\end{table}

\begin{table}[!h]
\begin{tabular}{|c|c|c|c|c|}
\hline
Quantity &	Value &	Standard unc. &	Sens. Coeff. &	Unc. contribution\\
\hline
\hline
$N_C$ &	$800$ &	$100$ &	$2 \ast 10^{-4}$ &	$3 \ast 10^{-2}$\\
\hline
$N_\xi$	 & $5000$ &	$1000$ &	$-2 \ast 10^{-5}$	 & $-2 \ast 10^{-2}$\\
\hline
$N_{BG}$	& $380$ & 	$30$ &	$-2 \ast 10^{-4}$ &	$-6 \ast 10^{-3}$\\
\hline
$\alpha_{exp}$ &	$0.079$ & & &			$0.009$\\
\hline
\end{tabular}
 \caption{Uncertainty Budget ($k$=2) associated with INRIM measurement(in joint measurement with PTB)}
 \label{tab:26inrim}
\end{table}

\begin{table}[!h]
\begin{tabular}{|c|c|c|c|c|}
\hline
Quantity &	Value &	Standard unc. &	Sens. Coeff. &	Unc. contribution\\
\hline
\hline
$N_C$ &	$900$ &	$70$ &	$2 \ast 10^{-4}$ &	$2 \ast 10^{-2}$\\
\hline
$N_\xi$	 & $5300$ &	$900$ &	$-2 \ast 10^{-5}$	 & $-2 \ast 10^{-2}$\\
\hline
$N_{BG}$	& $530$ & 	$40$ &	$-2 \ast 10^{-4}$ &	$-6 \ast 10^{-3}$\\
\hline
$\alpha_{exp}$ &	$0.076$ & & &			$0.007$\\
\hline
\end{tabular}
 \caption{Uncertainty Budget (k=2) associated with PTB measurement (in joint measurement with INRIM)}
 \label{tab:26ptb}
\end{table}

\begin{table}[!h]
\begin{tabular}{|c|c|c|}
\hline
& INRIM &	PTB \\
\hline
$\alpha_{exp}$ & 0.079 $\pm$ 0.009 & 0.076 $\pm$ 0.007\\
\hline
\hline
& INRIM &	NPL \\
\hline
$\alpha_{exp}$ & 0.065 $\pm$ 0.005 & 0.068 $\pm$ 0.005\\
\hline
\end{tabular}
 \caption{Summary of the results of the joint measurements performed by INRIM, NPL and PTB ($k$=2).}
 \label{tab:sum}
\end{table}

\section{Dependence on the coincidence window}
\label{app:w}
\begin{figure}[!h]
\centering
\includegraphics[scale=0.5]{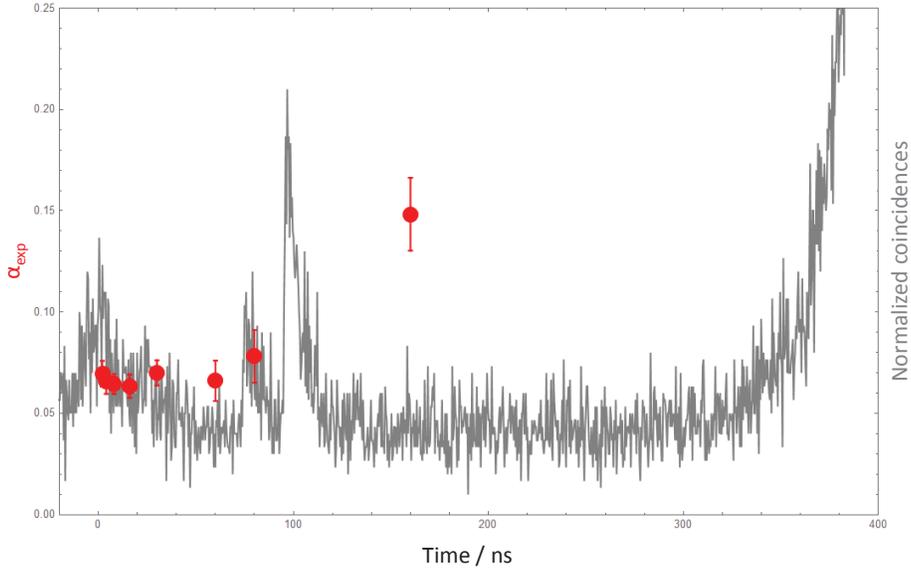}
\caption{Value of $\alpha_{exp}$ as a function the coincidence window $w$ (k=1). For the sake of clarity, the data (i. e., the red dots) are compared to the normalized chronogram of the coincidences (without background subtraction), shown in gray.}
\label{fig:finestra}
\end{figure}

To prove that the estimation of $\alpha$ is independent of the choice of the time interval of integration, we performed an analysis of the values of the measurand obtained by varying the coincidence window $w$. The results are shown in Fig. \ref{fig:finestra}, demonstrating that, as far as the backflash peaks are not included in the integration, the estimate is consistent independently of $w$.

\newpage

\section{conclusions}
A pilot study on the characterization of a pulsed-pumped test SPS based on a NV centre in nanodiamonds was performed by INRIM, NPL and PTB and hosted by INRIM.
 This study will greatly benefit the single-photon metrology community, as well as rapidly-growing quantum-technology-related industries. The main results of this study was the development of a standardized measurement technique as well as an uncertainty estimation procedure. The validity of the technique (system-independent  and unaffected by the non-ideality of the apparatus) is demonstrated by the results obtained, yielding for all the participants estimated values of $g^{(2)}(0)$ that are compatible within the uncertainty ($k$=2).
 
\section*{Acknowledgements}
This work was funded by: project EMPIR 14IND05 MIQC2, EMPIR 17FUN06 SIQUST, EMPIR 17FUN01 BECOME (the EMPIR initiative is co-funded by the European Union's Horizon 2020 research and innovation programme and the EMPIR Participating States); DIESIS project funded by the Italian National Institute of Nuclear Physics (INFN) - CSN5 within the "Young research grant" scheme; PiQUET project funded by the "Infra-P" program of the Regione Piemonte (POR-FESR 2014-2020 funding scheme);  "Departments of Excellence" (L. 232/2016) project funded by the Italian Ministry of Education, University and Research (MIUR).

\newpage

\appendix
\section{Lifetime estimation}
\begin{figure}[!h]
\centering
\includegraphics[scale=1.3]{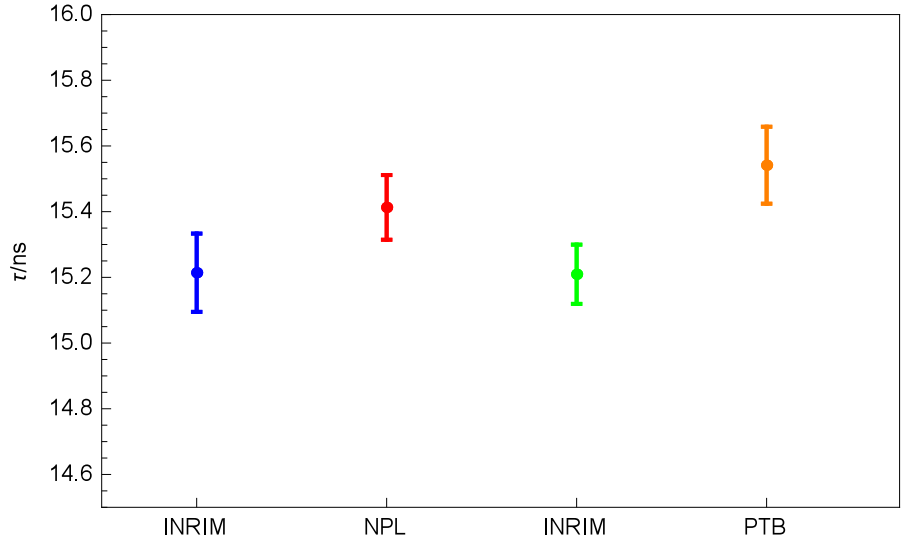}
\caption{Results of the emitter's lifetime $\tau$ (k=1) joint measurement respectively performed by: INRIM (blue dot) and NPL (red dot), INRIM (green dot) and PTB (orange dot).}
\label{fig:g2_summ_2}
\end{figure}

The mean lifetime associated with the source has been estimated by numerically fitting the coincidence histograms (as in Fig. 1) via the single-exponential function \cite{fit1,fit2,fit3}

\begin{equation}
f(\tau)=a+b\sum_{n=-\infty}^{+\infty}\Bigl( 1-\frac{\delta_{0n}}{c}\Bigr)e^{-\frac{|\tau-n\Delta t|}{d}},
\end{equation}

where $a$ corresponds to the number of background coincidences, $b$ is a normalization factor, $\delta_{0n}$ is the Dirac Delta, $c$ is the number of excited emitters, $n$ is the excitation pulse number, $\Delta t$ is the excitation period and, finally, $d$ accounts for the lifetime (convoluted with the detectors' jitter) of the center. Fig. \ref{fig:g2_summ_2} shows the results of the lifetime estimation independently performed by the partners. Each value in the plot represents the mean of the results of 10 fits (one for each experimental run performed by one partner).Averaging the results, it is obtained the value t$_{LIFE}$= (15.34 $\pm$ 0.08) ns. 

\end{document}